# Resource Discovery in Trilogy

## Abstract

Trilogy is a collaborative project whose key aim is the development of an integrated virtual laboratory to support research training within each institution and collaborative projects between the partners. In this paper, the architecture and underpinning platform of the system is described with particular emphasis being placed on the structure and the integration of the distributed database. A key element is the ontology that provides the multi-agent system with a conceptualisation specification of the domain; this ontology is explained, accompanied by a discussion how such a system is integrated and used within the virtual laboratory. Although in this paper, Telecommunications and in particular Broadband networks are used as exemplars, the underlying system principles are applicable to any domain where a combination of experimental and literature-based resources are required.

**Topics:**

    Architecture: agents, bus, distributed.

**Type of submission:**

    System paper

**Author Details:**

| **Mr Franck Chevalier** | **Dr. David Harle** | **Prof. Geoffrey Smith** |
|---|---|---|
| Dept. of EEE | Dept. of EEE, | Dept. of EEE, |
| University of Strathclyde | University of Strathclyde | University of Strathclyde |
| 204 George St, | 204 George St, | 204 George St, |
| Glasgow G1 1XW | Glasgow G1 1XW | Glasgow G1 1XW |
| United Kingdom | United Kingdom | United Kingdom |

**Email:** f.chevalier@eee.strath.ac.uk    d.harle@eee.strath.ac.uk    d.g.smith@eee.strath.ac.uk

**Tel:**    +44 (0)141-548-2090    +44 (0)141-548-1727    +44 (0)141-548-2537

**Fax:**    +44 (0)141-552-4968

# 1   Introduction

With the Internet explosion, it becomes increasingly difficult for researchers in various domains to find relevant information using traditional search engines. When they do, there are no means by which they can experiment or implement what they have just discovered. A virtual laboratory that would provide researchers with a virtual environment, where particular documents and experiments relevant to the users interests would be suggested, would be highly valuable. Tools to work on collaborative projects is also an area which generates a lot of interest in the computer science and librarian societies. Furthermore, making the expertise of different institutions, and the experience gained in previous projects available to a community of users would be a crucial step in setting up a truly distributed virtual laboratory.

Existing information retrieval systems (IRS) [1][2] do not offer such a service. They do not provide users with equipment to experiment and analyse given problems although instances of a Remote Experimentation Paradigm (REP) can be found on the Net. A REP, demonstrated by the Swiss Federal Institute of Technology (Lausanne), facilitates the remote control of a stepper motor and enables the user to observe the response of the motor to electrical impulses via the use of a video link and an HTTP server[3]. Although this technique brings a new dimension to the Internet, it suffers from several disadvantages

1. Should the stepper motor fail for any reason, the user does not know why the experiment does not work.

2. As there is only one physical motor and no scheduling, the system can only be used by one user at a time.

3. There is no online database associated with this experiment so the user must have some a-priori knowledge about the subject prior to accessing the system.

4. Collaborative work is not possible.

Trilogy is a virtual laboratory that addresses the above problems by dynamically linking experiments and virtual documentation to provide the user with a complete view of the area they are to specialise in. In effect, the document/experimentation paradigm is essential to maintain a complete picture of the area under study, specifically in the area of engineering. The virtual laboratory does not attempt to replicate the Internet by having non-exhaustive lists of documents, but provides a meaningful virtual environment where geographically separate institutions with complementary interests can interact and collaborate in a meaningful manner.

It may also be useful for the researcher to collaborate with other users in learning about a given tool. In addition to providing access to information and tools, and facilitating collaboration, experience from former projects and researchers can be made available to current users in the form of experiments as well as publications and reports. Additionally, the virtual laboratory can pro-actively provide a user, based on their profile and interests, with information, and on other users (with similar interests) queries.

# 2   System Architecture

Given the demands placed upon such a virtual laboratory coupled with its distributed nature, software agent technology is a natural choice to form the underlying system architecture

Before considering the system architecture in detail it is necessary to first define the terms "resource" and "service". A resource is the basic element within the laboratory and is used, via the intelligent agents, to provide the basic functionality. A resource can be a software tool such as a network simulator (for example OPNET®, NS® and hand-crafted C simulators), a display tool (e.g. Xgraph, PDF® viewer) or a mathematical tool (Matlab®). Crucially, a resource can also be experimental hardware that is to be accessed remotely by users of the laboratory. In the Trilogy laboratory such a resource takes the form of an ATM testbed containing 6 ATM switches and 2 cell blaster interfaces. In the context of the virtual laboratory, one of the most crucial resources is the distributed database. It is this part of the laboratory that is the focus of this paper.

A service is a resource as seen from the user's perspective. One resource can provide several services. For instance, the search by keyword and search by topic are two services that can be provided by the database resource. Alternatively, a single experimental resource can be used to provide a number of experiments when configured appropriately: each experiment represents a different service. Finally, a single service may make use of a variety of resources, e.g. a call to an a particular experiment may involve invoking a call to experimental hardware along with a database enquiry to provide the necessary documentation – tutorials, instructions or associated publications.

The virtual laboratory is a distributed, concurrent system that must be easily extensible as new tools or information resources are added. Agents are independent software entities that are able to act on their environment, e.g. an agent may perform a keyword search on the database it is managing, and interact with other agents in solving problems. The Trilogy Virtual Laboratory is designed as a multi-agent system in which agents play one of three distinct roles: the role of a personal assistant, a mediator or a resource provider[4][5]. Figure 1 illustrates the general architecture in terms of these key elements. The diagram indicates how such elements, which make up the agent layer, provide the linkage between the users and services represented by the laboratory resources.

The personal assistant agents (PAA) represent the users' interests within the system. These agents have several functionality's including storing the user's profile and history and performing some searches in the system on the behalf of "their" users.

The mediator agents provide a number of intermediate information services to agents within the Trilogy system. They act like the Yellow Pages and have a specialised knowledge of the domain to present an abstraction about that information. For instance, they know what resource provides what service and what database contains what type of information. These agents play an active role between the personal assistant agent layer and the resource agent layer.

The last type of agent is the resource agent (RA). The main function of this type of agent is to manage the resource it is associated with. Additional roles include scheduling experiments for the users when a resource can only have one instance of its service at a time, letting the user know the reason of a resource failure when it "crashes", and advertising the services provided by its resource to one or several mediator agents.

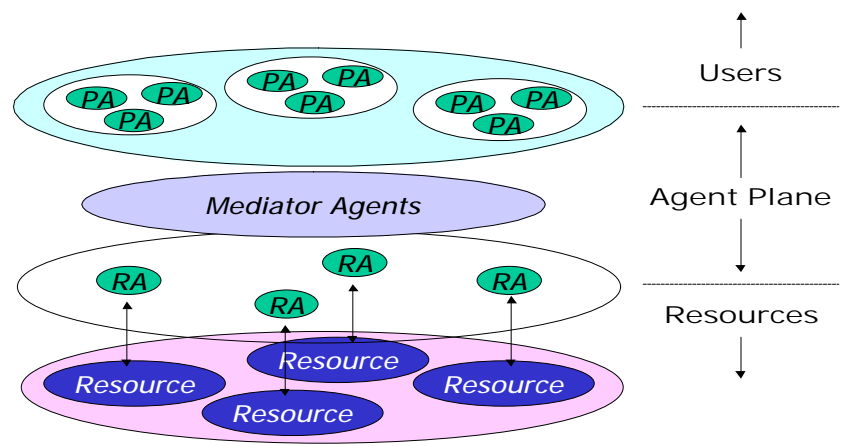

**Figure 1**: System Architecture

# 3 Trilogy Database

## 3.1 Information Retrieval System Selection Criteria

One of the crucial resources of the laboratory is the database. The challenges it presents are now examined.

### Cataloguing Standard and Automation

It was important to use a flexible system that can index different type of documents (e.g. HTML, Postscript, nested types and customised formats). In effect, the database stores conventional information sources but is also used to store information about the available resources from the laboratory. Examples of such documents are the accurate descriptions of the OPNET experiments that are available from the Trilogy common model library. Perhaps, the most widely used indexing standard is the USMARC standard that has been designed following an agreed set of principles known as the Anglo-American Cataloguing Rules (AACR). Such a standard is impractical for the case of Trilogy as it requires an expert librarian to manually build such a record for each document. Therefore a much simpler indexing scheme, that could be automatically created from the any documents is required.

### Collection Maintenance

It is a very common experience for a WWW user to find an interesting URL, to bookmark it, and, when attempting to access it again some time later, to get the infamous "404 error". It is important to identify the dynamic nature of the resources on the Web: they can change content, location or just disappear. In effect, a user does not want to be in a position where the result of a query only points to non-existent documents. An automated update of the catalogue is therefore required on a regular basis to maintain consistent records.

### Scalability and repositories distribution

Scalability is critical as the volume of documents will increase as the number of users in the system increases; users may want to add different references to the database. Furthermore, the database must be distributed over the collaborative sites; as centralised information retrieval system is relatively inefficient for large collections and it is much more efficient to have smaller repositories and to logically group information by topic.

The Harvest information retrieval system has been identified to be one candidate that provides the required features.

## 3.2 Harvest System

Harvest is a distributed system architecture that supports object (such as files, packages, HTML documents) location on the Internet, and can inter-operate with WWW clients and with HTTP, FTP, Gopher and Netnews . The Harvest architecture consists of four parts: the Gatherer, the Broker, the Object Cache and the Replication Manager[6].

The Gatherer summarises information at archive sites and creates object summaries called Summary Object Interchanged Format objects (SOIF). The Broker provides the indexing and the query interface to the gathered information. Brokers retrieve information from one or more gatherers or other Brokers, and incrementally update their indexes in an automated manner [6].

To decrease the network load, the replication Manager maintains several identical copies of a Broker: it allows a number of different Brokers to be run over the same data set to split the load between two different machines for heavily used brokers. Finally, the Object cache maintains local copies of popular objects to improve access performance.

### 3.3 Database Integration

Details about the location of particular piece of information, and the protocol used to retrieve that information must be completely transparent to the user [5]. The architecture illustrated in Figure 2 allows this transparency as the user can only see the graphical interface

*Repositories logical arrangement*

As shown in Figure 2, the different topic repositories (Brokers) are distributed over a number of different sites. The repositories are logically arranged according to well defined topics specified in an ontology. The ontology and the technique that is used to classify the documents are further discussed in Section 4.

This arrangement is very flexible as it allows new sites to be added to the virtual laboratory in a straight forward manner; a site with expertise in a particular topic can have their own personal Harvest broker which can be managed by a resource agent situated in a different location. The only requirement is to a one-to-one mapping between the resources and the resource agents.

Finding the appropriate database for a given topic is a crucial issue. This is achieved by the use of the mediator agent specialised in literature search. This agent will provide the user agent with the name of the resource agent to contact based on its knowledge of the domain.

For instance, if a user is looking for information on "ATM connection admission control", the Mediator Agents indicates to the PAA the location of the resources (topic brokers) that will provide this information. This is achieved by indicating the name of the resource (which is uniquely defined in the system) so that the personal assistant agent (PAA) can subsequently contact the appropriate resources and hence, answer the initial query.

*Integration*

The resource agents are generic and can manage any resources in the system: the same agent is used to manage a software tool, the ATM testbed or the information repositories (Broker). In Figure 2, it can be observed that a software interface is required between the RA and the resource it manages; the broker in this case. This extra layer allows the broker and its resource agent (RA) to communicate. The broker informs the RA of the services that it can provide, the number of inputs required and the number of instances that are allowed to be run concurrently. The latter prevents the system from being overloaded as many users may want to access the same service at the same time.

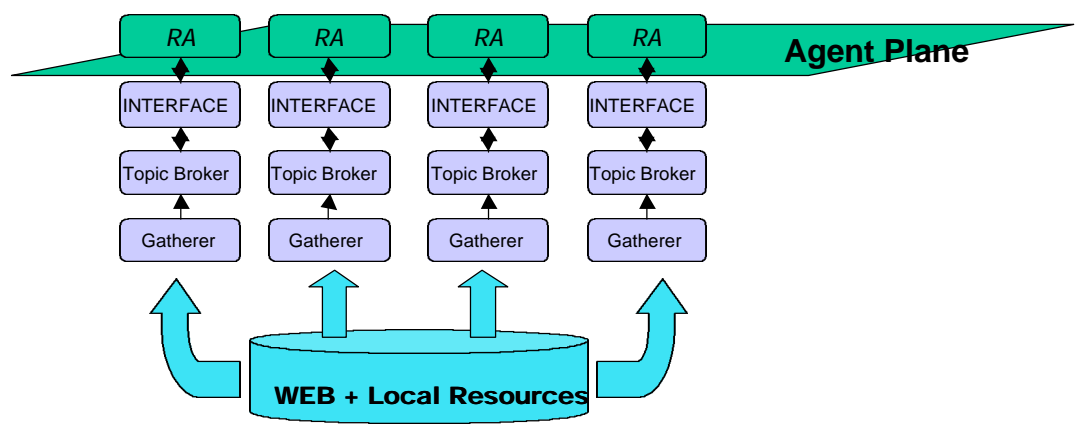

**Figure 2**: Database Architecture

If more than one user requires access to a resource, scheduling is possible whereby requests are queued and the appropriate PA is informed.

In the same way, the broker can understand the content of the messages received from the RA such as the query string, the user name (to store results in the user's directory). Depending on the resource, this interface may be different as different functionalities are required for different resources. Extensive description of other resources can be found in [5] and are outwith the scope of this paper.

*Database population*

An important feature of Trilogy is the information and experience re-usability. All researchers must be able to contribute to the database population using simple means. The interface that has been developed for this process is both simple and effective. This interface is dynamically generated depending upon the type of documents to be added, these include: conference articles, books, book chapters, journal articles, thesis and technical report. This interface is shown on Figure 3 for the special case of conference articles.

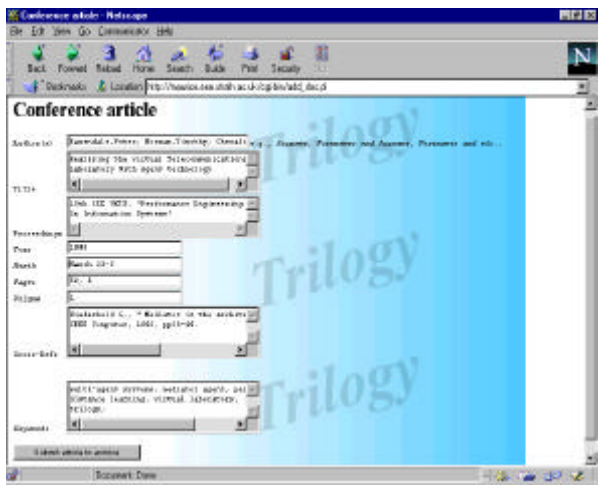

**Figure 3**: Add Document Interface

The advantages of this architecture are numerous. First, this system is highly scalable as new brokers specialised in different topics can be added easily to the system. The only integration required is the specification of keywords and topics that the new repository is specialised in at the resource interface level. The generic RA that is associated with the new broker will automatically advertise its services to the literature-search Mediator Agent. Secondly, the retrieval time decreases as the individual database to be searched are significantly smaller. Finally, this allows a controlled level of replication in the different databases as a document can be linked with several topics defined in the ontology making this system even more robust to individual site failure.

The key requirement associated with this architecture is the need for an ontology which is specific to the domain of application (Broadband Networks in our case). This ontology is used to allow the domain information to be encoded precisely. That is defining the legal relations between entities and relate individual entities to classes. In the context of the database, the ontology provides a mechanism for expressing queries beyond the keyword based approach.

# 4 Trilogy Ontology

## *4.1 Ontology description*

The word "ontology" seems to generate a lot of controversy in the AI world. In the context of knowledge sharing, it can be defined as *a specification of a conceptualisation*[7]. That is, an ontology is a description of the concepts and relationship that can exist in the domain of application.

Although it is recognised that ontologies are both difficult to develop and to apply, they increasingly become a subsidiary to more traditional institutional structure, for organising quantity of information that increases in a three dimensional way – volume, concepts and formats. In [2], advanced ontologies are used to modeling the content, the services and the licence agreements of the **Beethoven** intelligent information retrieval system. In the context of the Beethoven project, the University of Michigan Digital Library (UMDL) Ontology group has developed a formal conceptualisation of bibliographic relations. The entities that populate the database are defined in terms of CONCEPTIONs (associated with the idea), EXPRESSIONs (the work's content), MANIFESTATIONs (publishing format), DIGITISATIONs (encoding) and INSTANCEs (specific copy). This knowledge base involves mapping data from the MARC records to the ontology and reasoning about the data to establish relationship[8].

However, differences exist between the Trilogy ontology and the ontology described above. Firstly, by definition, the domain of application of the virtual laboratory is restricted to Broadband Networks. Therefore the ontology must be concise to describe the domain adequately. Secondly, in [2], the conceptualisation of the database relies on the mapping of MARC records. In Section 2, the drawbacks associated with such a catalogue format have been discussed.

The ontology architecture and the different relationships that exist between the different concepts are now explained. In the Trilogy ontology, two level of abstraction exist: the **concept** (or topic) and the **keyword**, where a document concept is defined in terms of keywords it is associated with. Within the topic paradigm, a *parent-child* relationship exits. For instance, the concept "Wireless ATM" is a child of the concept "ATM General". To characterise the keyword - concept relationship, weights are used. In effect, it is essential to be able to quantify this relationship as a keyword may be linked to two different concepts, but with different emphasis.

The weight is an integer that is bounded between 1 and 20. This relationship as well as the weight associated with it, is dynamic ion nature : it must reflect the expertise and the current research that is being carried out at the collaborating sites. A web interface to the system ontology allows new concepts to be added and relative weighting to be adjusted. The concept-keyword relationship is partly in Figure 4.

In the concept hierarchy, the first level (main) topics include ATM general, SDH General, Analytical Models, Simulation Models, High Speed LANs and MANs, and Optical Networks.

An essential difference between this architecture and the traditional hierarchical classification (such as the Dewey system) is that documents or experiments can be described in terms of several topics with different emphasis. Let us consider a paper named: "Cell Level Simulation in ATM Networks". Intuitively, two different concepts can be extracted from this title: ATM and Simulation. The user may be familiar with ATM but not with the different simulation techniques that are employed to simulate ATM network's behaviour. Thus, the PAA may perform a literature review on simulation techniques on the behalf of the researcher.

### 4.2 Ontology Implementation and Integration

The way the ontology is used within the system is now described. In each Harvest broker, a number of summary object files (SOIF) reside. It is relatively straight forward to produce an index of the keywords contained in each of these SOIFs. This index is then mapped to concepts by looking up the keyword-concept table. The weight associated with each keyword is summed up and normalised to the document. The last step is to produce an index (inverted) of the concepts associated with theirs SOIF and weight/emphasis. The overall process is illustrated in Figure 5.

| Keyword | Concept | Weight |
| --- | --- | --- |
| Aal | Adaptation Layer And Transport Layer | 20 |
| Aal | ATM Introduction | 3 |
| Connection admission control | ATM Bandwidth allocation | 10 |
| Regenerator section | SDH Networking and Components | 8 |
| WDM | Wavelength Division Multiplexing | 15 |

**Figure 4**: Keyword Concept Table

These tables are used by the agents for various purposes such as finding a resource that corresponds to the users requirement, finding another user whose interests are related or just finding documents about a particular experiment available from the Trilogy Laboratory.

This indexing process is repeated on a regular basis to keep a consistent index of the files that are indexed in the database. This operation is performed during the week-end and requires an average of one hour per broker (~3000 documents) on a half loaded Sparc 4 station.

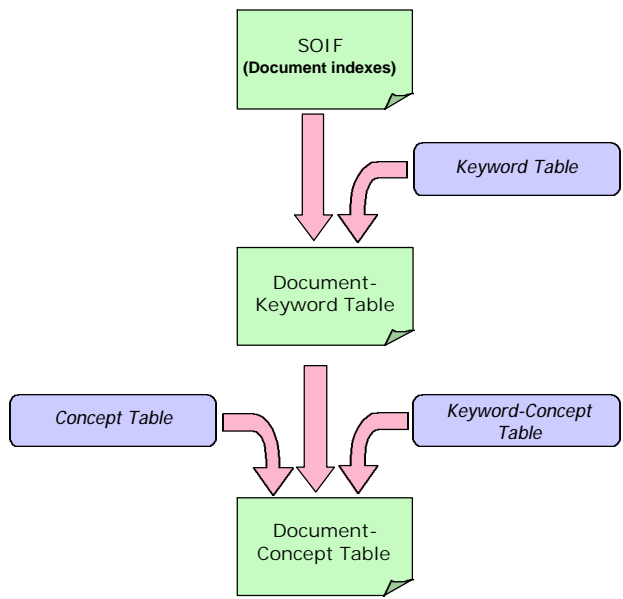

**Figure 5**: Indexing Process

## 5 Conclusion

The term virtual laboratory suggests more than just a static information retrieval system. This paper has introduced some of the challenges that were associated with the deployment of a distributed database in a multi-agent environment.

The importance of the three dimensional proliferation of information and experiment has been underlined since more documents, more users, and ultimately more sites will be added to the laboratory. The benefits of managing resources by the community of agents has been clearly shown .

The use and the integration of an ontology has also been discussed and its advantages, in terms of providing new services to the users and coping with a continually evolving topic of research, has been addressed.. It has been shown that the ontology provides the user with functionalities exceeding those provided by more traditional IRS.

The IRS architecture implemented in the Trilogy project offers a unique way for novice and experienced researchers to either work individually or to collaborate in projects. This has been borne out by a number of projects spawned from this work[9].

# 7 Acknowledgement

The authors would like to acknowledge the support for this work provided by the EPSRC for the Trilogy project (Grant no GR/L09714) and the Trilogy Members:

**Queen Mary Westfield University**
Norman T., Jennings N, Culthbert L

**Durham University**
Baxendale P., Mellor J., Mars P.